\documentclass[journal]{IEEEtran}
\usepackage{graphicx}
\usepackage{amsmath}
\usepackage[colorinlistoftodos]{todonotes}
\usepackage{subfigure}
\usepackage{multirow}
\usepackage{algorithm}
\usepackage{algorithmic}
\usepackage{pdfpages}
\usepackage{url}

\ifCLASSINFOpdf
\else
\fi

\begin{document}
	
	\title{Agent with Warm Start and Adaptive Dynamic Termination for Plane Localization in 3D Ultrasound}
	
	\author
	{Xin Yang, Haoran Dou, Ruobing Huang, Wufeng Xue, Yuhao Huang, Jikuan Qian, Yuanji Zhang, Huanjia Luo, Huizhi Guo, Tianfu Wang, Yi Xiong, Dong Ni
		
		\thanks{
			© 2021 IEEE.  Personal use of this material is permitted.  Permission from IEEE must be obtained for all other uses, in any current or future media, including reprinting/republishing this material for advertising or promotional purposes, creating new collective works, for resale or redistribution to servers or lists, or reuse of any copyrighted component of this work in other works.
			
			This study was supported in part by the National Key R\&D Program of China (No. 2019YFC0118300); Shenzhen Peacock Plan (No. KQTD2016053112051497, KQJSCX20180328095606003); Natural Science Foundation of Shenzhen (JCYJ20190808115419619).
			
			Xin Yang, Haoran Dou, Ruobing Huang, Wufeng Xue, Yuhao Huang, Jikuan Qian, Tianfu Wang and Dong Ni are with the National-Regional Key Technology Engineering Laboratory for Medical Ultrasound, Guangdong Key Laboratory for Biomedical Measurements and Ultrasound Imaging, School of Biomedical Engineering, Health Science Center, Shenzhen University, Shenzhen, China, and also with the Medical UltraSound Image Computing (MUSIC) Lab, Shenzhen, China (email:yangxinknow@gmail.com, douhaoran2017@email.szu.edu.cn)
			
			Yuanji Zhang, Huanjia Luo, Huizhi Guo, Yi Xiong are with the Department of Ultrasound, Shenzhen Luohu People's Hospital, the Third Affiliated Hospital of Shenzhen University, Shenzhen, China
			
			Xin Yang and Haoran Dou contribute to this work equally.
			
			Corresponding Authors: Yi Xiong (email: 13352995536@163.com), Dong Ni (email: nidong@szu.edu.cn)
			
			The code for the techniques presented in this study can be found at: https://github.com/wulalago/AgentSPL
		}
	}
	
	\maketitle
	
	% ABSTRACT START
	\begin{abstract}
		Accurate standard plane (SP) localization is the fundamental step for prenatal ultrasound (US) diagnosis. Typically, dozens of US SPs are collected to determine the clinical diagnosis. 2D US has to perform scanning for each SP, which is time-consuming and operator-dependent. While 3D US containing multiple SPs in one shot has the inherent advantages of less user-dependency and more efficiency. Automatically locating SP in 3D US is very challenging due to the huge search space and large fetal posture variations. Our previous study proposed a deep reinforcement learning (RL) framework with an alignment module and active termination to localize SPs in 3D US automatically. However, termination of agent search in RL is important and affects the practical deployment. In this study, we enhance our previous RL framework with a newly designed adaptive dynamic termination to enable an early stop for the agent searching, saving at most $67\%$ inference time, thus boosting the accuracy and efficiency of the RL framework at the same time. Besides, we validate the effectiveness and generalizability of our algorithm extensively on our in-house multi-organ datasets containing 433 fetal brain volumes, 519 fetal abdomen volumes, and 683 uterus volumes. Our approach achieves localization error of $2.52mm/10.26^\circ$, $2.48mm/10.39^\circ$, $2.02mm/10.48^\circ$, $2.00mm/14.57^\circ$, $2.61mm/9.71^\circ$, $3.09mm/9.58^\circ$, $1.49mm/7.54^\circ$ for the transcerebellar, transventricular, transthalamic planes in fetal brain, abdominal plane in fetal abdomen, and mid-sagittal, transverse and coronal planes in uterus, respectively. Experimental results show that our method is general and has the potential to improve the efficiency and standardization of US scanning. 
	\end{abstract}
	% ABSTRACT END
	
	\IEEEpeerreviewmaketitle
	
	% INTRODUCTION START
	\section{Introduction}
	
	\IEEEPARstart{U}{ltrasound (US)} is the primary screening method for assessing fetal health and development due to its advantages of being low-cost, real-time and radiation-free~\cite{None2013ISUOG}. Generally, during the diagnosis, sonographers first scan pregnant women to obtain US videos or volumes, then manually localize standard planes (SPs) from them. Next, they measure biometrics on the planes and make the diagnosis~\cite{SalomonPractice}. Of these, SP acquisition is vital for subsequent biometric measurement and diagnosis. However, it is very time-consuming to acquire nearly thirty SPs during the diagnosis and the process often requires extensive experiences due to the large difference in fetal posture and the complexity of SP definitions. Thus, automatic SP localization is highly expected to improve the diagnostic efficiency and decrease operator-dependency.
	
	\subsection{Standard Plane Localization}
	
	2D and 3D US are two typical modalities used in prenatal diagnosis. 2D US is easy to use and has better imaging quality. However, automatic 2D SPs localization may fail in detecting SPs when they are not scanned by clinicians due to the invisibility of the fetus and fine position of each plane.
	3D US can contain multiple SPs in just a single shot and has the inherent advantages of less user-dependency and more efficiency compared with 2D US~\cite{Namburete2014DiagnosticPE}. 
	Usually, after obtaining the 3D US, the sonographer shifts and rotates the current view plane to approach the SP. However, it is very challenging to manually localize SPs in the volume due to the huge search space, the large fetal posture variability and the low image quality. Therefore, the development of automatic methods for localizing SPs in 3D US would improve diagnostic efficiency and decrease operator-dependency by providing a de-specialized scanning method for non-experts.
	
	\subsection{Termination Strategy for Reinforcement Learning}
	In reinforcement learning (RL), a decision needs to be made by the agent as to whether to terminate the inference. The termination conditions are usually pre-set, such as reaching to the destination in \textit{MountainCar}~\cite{moore1990efficient}, pole's falling up in \textit{CartPole}~\cite{barto1983neuronlike}, etc. However, the termination conditions are often indistinct and can not be precisely determined in many tasks (e.g., object detection~\cite{caicedo2015active}, landmark detection~\cite{Ghesu2019MultiScaleDR}, SP localization~\cite{Dou10.1007/978-3-030-32254-0_33}, etc.). Specifically, in the SP localization, the agent might fail to catch the target SP and continue to explore without termination condition. One solution was to extend the action space with a further terminate action~\cite{caicedo2015active}. However, enlarging the action space will result in insufficient training. Several works terminated the agent searching by detecting oscillation~\cite{Ghesu2019MultiScaleDR} or the lowest Q-value~\cite{Alansary10.1007/978-3-030-00928-1_32}. Although no additional action was introduced, these approaches still required the agent to complete inference with maximum step, which is inefficient. Therefore, a dynamic termination strategy to ensure the efficacy and efficiency of the SP localization is highly desirable in the SP localization task.
	
	\subsection{Related Work}
	In our review of the related work on SP localization, we first introduce the approaches based on 2D US, and then we summarize the 3D US methods. Finally, we involve our previous deep RL-based algorithm.
	
	\subsubsection{Standard Plane Detection in 2D US}
	The early works~\cite{NI20142728,Yang6868086,Lei6867815,zhangPMID:22894427} selected SPs based on conventional machine learning methods (i.e., adaboost, random forest, support vector machine) through detecting key anatomical structures or landmarks of each frame in the video. Recent approaches made use of the convolutional neural network (CNN) due to its powerful ability in automatically learning hierarchical representations. The first two studies~\cite{Chen7090943, Chen10.1007/978-3-319-10581-9_16} built the CNN model with transfer learning technology to detect fetal SPs. Chen~\textit{et al.}~\cite{Chen7890445} then equipped the CNN with recurrent neural network (RNN) to capture the spatial-temporal information to detect three fetal SPs. Similar design can also be found in~\cite{Huang2017TemporalHT, Gao2017DetectionAC}. Baumgartner~\textit{et al.}~\cite{Baumgartner7974824} further proposed a weakly-supervised approach to detect 13 fetal SPs and locate region of interest in each plane. Inspired by~\cite{Baumgartner7974824}, Schlemper~\textit{et al.}~\cite{SCHLEMPER2019197} incorporated the gated attention mechanism into the CNN to contextualize local information for detecting SPs. More recently, some works~\cite{Wu7875138,Luo2019AutomaticQA,LIN2019101548} proposed to assess US image quality automatically. Wu~\textit{et al.}~\cite{Wu7875138} first introduced the quality assessment system of fetal abdominal plane by a cascade CNN. Luo~\textit{et al.}~\cite{Luo2019AutomaticQA} and Lin~\textit{et al.}~\cite{LIN2019101548} then proposed to assess the quality of fetal brain, abdomen and heart SPs by multi-task learning. These above methods showed the efficacy of detecting SPs and assessing image quality by transfer learning, spatial-temporal information, attention mechanisms and multi-task learning. However, automatic SP detection in 2D US still suffers from the high dependence on clinicians' scanning skills.
	
	\subsubsection{Standard Plane Localization in 3D US}
	Different from 2D US, localizing SPs in 3D US usually faces challenges of low image quality, large data size and huge search space. A number of works~\cite{Zhu10.1007/978-3-319-61188-4_13, NIE2017286,Lorenz10.1117/12.2292729,Chykeyuk10.1007/978-3-319-05530-5_6} formulated this task as a cascade pipeline (i.e. from landmark detection to SP regression) based on conventional machine learning methods. Although effective by using prior anatomical knowledge, the performance of these methods is still limited by landmark detection accuracy and testing case-model difference. Recently, Ryou~\textit{et al.}~\cite{Ryou10.1007/978-3-319-47157-0_24} proposed to locate the fetus by random forest and detect SPs by CNNs sequentially. Schmidt-Richberg~\textit{et al.}~\cite{Alexander10.1117/12.2512697} introduced a deep learning based regression framework to estimate SP locations. Li \textit{et al.}~\cite{Li10.1007/978-3-030-00928-1_45} proposed a deep neural network to move the estimated plane to the target SP iteratively. They further customized a RL-based agent for view plane searching in MRI volumes~\cite{Alansary10.1007/978-3-030-00928-1_32}. RL is promising for SP localization in 3D US due to its ability of mimicking experts' operation and exploring inter-plane dependency by the agent-environment interaction. However, the RL solution may suffer from its random initialization and empirical termination when its environment, such as the US volume, has strong noise, artifacts and large appearance variations.
	
	To address the issues mentioned above, our previous study~\cite{Dou10.1007/978-3-030-32254-0_33} proposed a RL based framework to automatically localize SPs in 3D US. We equipped the RL framework with a landmark-aware alignment module for warm start to ensure its effectiveness. In this module, we leveraged the CNN to detect anatomical landmarks in the US volume and registered them to a plane-specific atlas, thus providing strong spatial bounds and effective initialization for the RL. Furthermore, instead of passively and empirically terminating the agent inference, we introduced a learning-based strategy for active termination of the agent's interaction procedure through an RNN module. The learning-based strategy can achieve optimal termination adaptively, thus improving the accuracy and efficiency of the localization system.
	
	\begin{figure}[!htbp]
		\centering
		\includegraphics[width=1.0\linewidth]{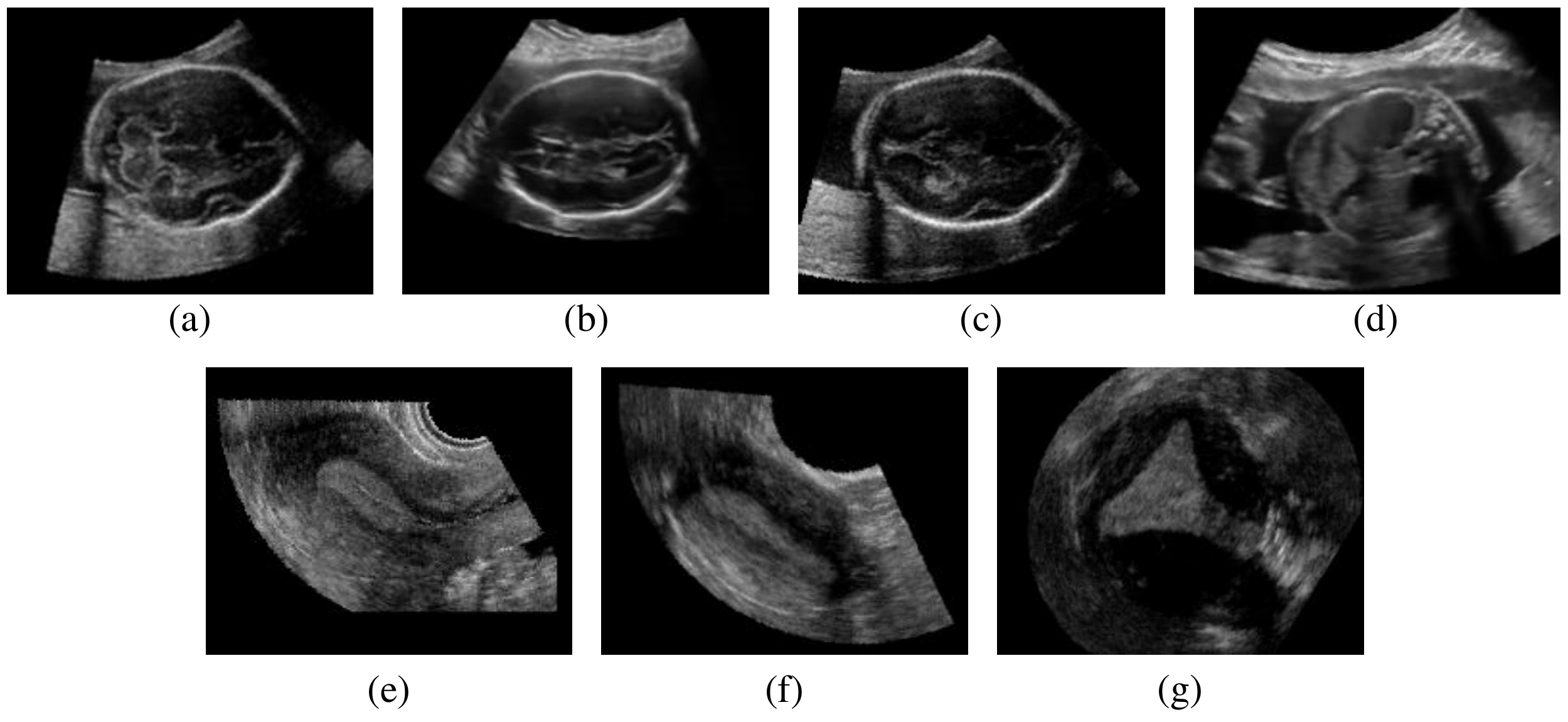}
		\caption{
				Sample cases of the studied seven target planes. (a)-(g) correspond to the transcerebellar, transventricular, transthalamic, abdominal, mid-sagittal, transverse, and coronal planes, respectively. Particularly, planes (a)-(c) are in the fetal brain; plane (c) is in the fetal abdomen; and planes (d)-(g) are in the uterus.
		}
		\label{fig:target_plane}
	\end{figure}
	
	\subsection{Contribution}
	In this study, we further improve the stableness, robustness and efficiency of our previous method~\cite{Dou10.1007/978-3-030-32254-0_33}. This article has considerable difference compared with the previous conference paper, which consists of:
	
	\begin{itemize}
		\item We design an adaptive dynamic termination based on our previous work~\cite{Dou10.1007/978-3-030-32254-0_33}, which enables an early stop for the agent searching, resulting in efficiency-steered localization system. Dynamic termination is an important yet unsolved problem in reinforcement learning; Our work provides the first effective solution for this and can be generalized to other similar scenarios.
		
		\item We validate the effectiveness and the generalizability of our method on a large multi-organ dataset including 433 fetal brain volumes, 519 fetal abdomen volumes, and 683 uterus volumes. Specifically, we propose to localize seven SPs (Fig.~\ref{fig:target_plane}) from multiple organs, in contrast to the two SPs from one organ~\cite{Dou10.1007/978-3-030-32254-0_33}. 
		
		\item We have conducted comprehensive experiments to validate the superiority of our method over existing ones in aspects of SP localization performance,  performance comparison,  computation efficiency, effectiveness of the proposed adaptive termination module, and biometric and qualitative evaluation of the obtained results from the aspects of clinical practice. 
	\end{itemize}
	
	% INTRODUCTION END
	
	% METHOD START
	\section{Method}
	
	\begin{figure*}[!htbp]
		\centering
		\includegraphics[width=0.9\linewidth]{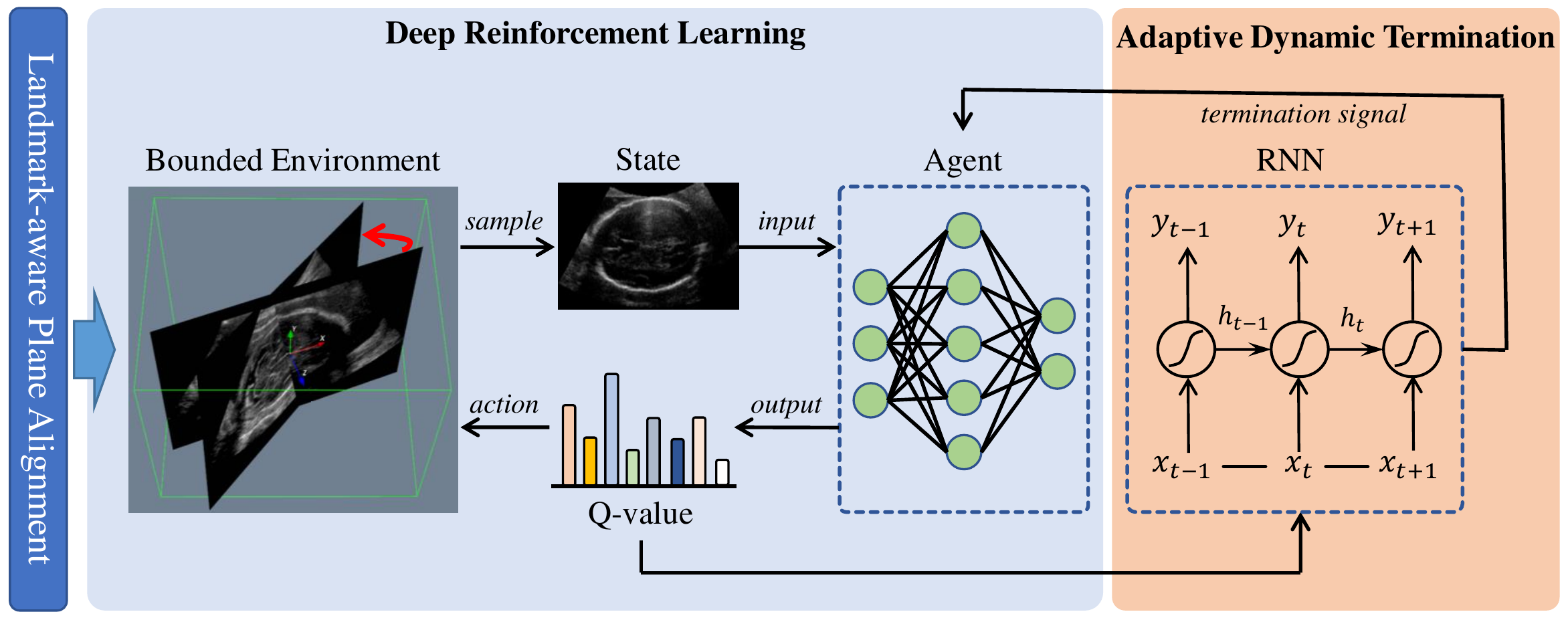}	\caption{Schematic view of our proposed framework. Our framework contains three modules: 1) the SP localization based on deep RL (middle), 2) the landmark-aware alignment to provide a warm start for efficient agent searching (left), and 3) the learning-based adaptive termination to improve the localization efficiency and accuracy (right).}
		\label{fig:framework}
	\end{figure*}
	
	As shown in Fig~\ref{fig:framework}, an automatic plane localization system for 3D US is proposed to imitate the diagnosis of experienced physicians. This system is implemented with a two-step unified RL framework.
	First, a landmark-aware alignment module~\cite{Dou10.1007/978-3-030-32254-0_33} is adopted to reduce large search space caused by the complex intrauterine environment and diverse fetal postures. 
	Then, a deep reinforcement model searches the target SPs within the bounded environment resulted from the alignment module. 
	An adaptive RNN-based termination module is adopted to dynamically stop the RL agent at the optimal interaction.
	
	\subsection{Deep Reinforcement Learning Framework}
	In the classical deep RL framework, the agent interacts with the environment $\mathcal{E}$ by making successive actions $a \in \mathcal{A}$ to maximize the expectation of reward, where $\mathcal{A}$ is the action space. Meanwhile, a plane in 3D space is modelled as $cos(\alpha)x+cos(\beta)y +cos(\gamma)z=d$, where $n=(cos(\alpha), cos(\beta), cos(\gamma))$ denotes the unit normal vector of the plane, and $d$ is its Euclidean distance from the origin. In this work, the origin is set as the center of an US volume. We therefore define the main elements of this plane-localization RL framework as follows:
	
	\textbf{State}: The state is defined as the reconstructed 2D US image from the volume given the current plane parameters. Since the reconstructed image size may change, we pad the image to a square and resize it to 224 $\times$ 224. In addition, we concatenate the two images obtained from the previous two iterations with the current plane to enrich the state information, which is similar to~\cite{Mnih2015HumanlevelCT}.
	
	\textbf{Action}: The action is defined as incremental adjustment to the plane parameters. The complete action space is defined as $\mathcal{A}= \left\lbrace \pm a_{\alpha},\pm a_{\beta},\pm a_{\gamma}, \pm a_{d}\right\rbrace $. Given an action, the plane parameters are modified accordingly (e.g. $\alpha_{i}=\alpha_{i-1} + a_{\alpha}$). We perform one action to adjust only one plane parameter with the others unchanged for each iteration. Specifically, the step size of angle adjustment is $a_{\alpha}=a_{\beta}=a_{\gamma}=1^\circ$, while the distance step size $a_{d}$ is set as 0.5 voxel in each iteration. 
	
	\textbf{Reward}: The reward signal defines the goal in a RL problem. It instructs the agent what policy should be taken to select the proper action. In this study, the reward is defined as whether the agent approaches or moves away from the target, which can be obtained by $r = sgn(\|P_{i-1}-P_{g}\|_2-\|P_{i}-P_{g}\|_2)$, where $P_{i}$, $P_{g}$ indicate the plane parameters of the predicted plane and the ground truth in iteration $i$, $sgn(\cdot)$ is the sign function. The universal set of the calculated reward signal is: $\lbrace +1,0,-1\rbrace $, where $+1$ and $-1$ indicate the positive and negative movement, respectively, and $0$ refers to no adjustment.
	
	\begin{figure}[!htbp]
		\centering
		\includegraphics[width=1\linewidth]{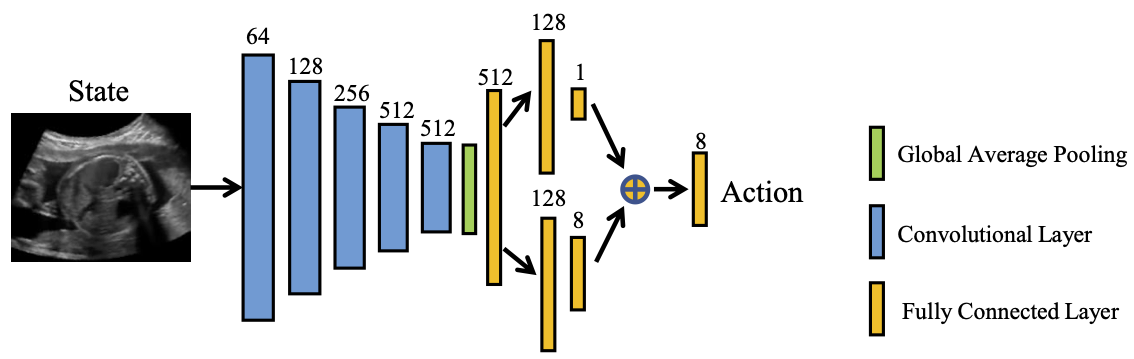}	
		\caption{Architecture of our agent neural network. Blue, green and yellow rectangles represent convolutional layers global average pooling and fully connected layers, respectively.}
		\label{fig:dueldqn}
	\end{figure}
	
	\textbf{Agent}: The agent is a policy component that outputs the action via interacting with environment. In this study, we adopt the Q-learning~\cite{watkins1992q} as the solution for the SP localization. Different from the existing work using a deep neural network to estimate the Q-value directly~\cite{Mnih2015HumanlevelCT}, the dueling learning~\cite{Wang2016DuelingNA} is utilized to encourage the agent to learn which states should be weighted more and which are redundant in choosing proper actions. Specifically, the Q-value function is decomposed into a state value function and a state-dependent action advantage function, respectively.
	
	As shown in Fig.~\ref{fig:dueldqn}, the deep duel neural network takes the state as input and outputs the action. In this work, we use the pre-trained VGG~\cite{Simonyan2015VeryDC} as the convolutional backbone. The number of features in each layer is 64, 128, 256, 512, 512, respectively. To mitigate the gradient vanishing issue, we add batch normalization layer~\cite{Ioffe2015BatchNA} after each convolutional layer in the neural network. The extracted high-level features are then fed into two independent streams of fully connected layers to estimate the state value and the state-dependent action advantage value. The hidden units of fully connected layers are 512, 128, 1 in the state value estimation stream, and 512, 128, 8 in the state-dependent action value estimation stream, respectively. The outputs of the two streams are fused to predict the final Q-value.
	
	\textbf{Replay Buffer}: 
	The replay buffer is a memory container that stores the transitions of the agent to perform experience replay for learning procedure. Element $transition$ is typically represented with a vector $(s_t, a_t, r_t, s_{t+1})$, where $s_{t}, a_{t}, r_{t}$ denote the state, action and reward at the step $t$. In this study, the prioritized replay buffer~\cite{schaul2015prioritized} is adopted to improve the learning efficiency.
	
	\textbf{Training Loss}: As explained above, we decompose the Q-value function, $Q(s,a;w)$, into two separate estimators including the state value function $V(s;w_{c},w_{v})$ and the state-dependent action advantage function $A(s, a; w_{c}, w_{s})$, where $s$ is the input state of the agent, $a$ is the action, $w_{c} ,w_{v}, w_{s}$ represent the parameters of the convolution layers and the two streams of fully-connected layers, respectively, and $w=\{ w_{c}, w_{v}, w_{s} \}$. The Q-value function of the agent is calculated as:
	\begin{equation}
		\begin{aligned}
			\label{eqa:Q}
			Q(s,a;w) = &V(s;w_{c},w_{v})+A(s,a;w_{c},w_{s})\\
			&- \frac{1}{|A|}\sum_{a} A(s,a;w_{c},w_{s})
		\end{aligned}
	\end{equation}
	where $\left|A\right|=8$ denotes the size of the action space. The loss function for our framework is then defined as:
	\begin{equation}
			\begin{aligned}
				\label{eq:loss}
				L(w) = E_{s,a,r,\hat{s} \sim U(M)}&[(r + \gamma\underset{\hat{a}}{max}Q_{target} (\hat{s},Q(\hat{s},\hat{a};w);\tilde{w})\\
				&-Q(s,a;w))^{2}]
			\end{aligned}
	\end{equation}
	where $\gamma$ is a discount factor to weight future rewards; $\hat{s}$ and $\hat{a}$ are the state and the action in next step; $U(M)$ represents uniform data sampling from the experience replay memory $M$; $w$ and $\tilde{w}$ are the parameters of Q network ($Q(w)$) and target Q network ($Q_{target}(\tilde{w})$).
	
	\subsection{Landmark-aware Plane Alignment for Warm Start}
	\begin{figure}[!htbp]
		\centering
		\includegraphics[width=0.95\linewidth]{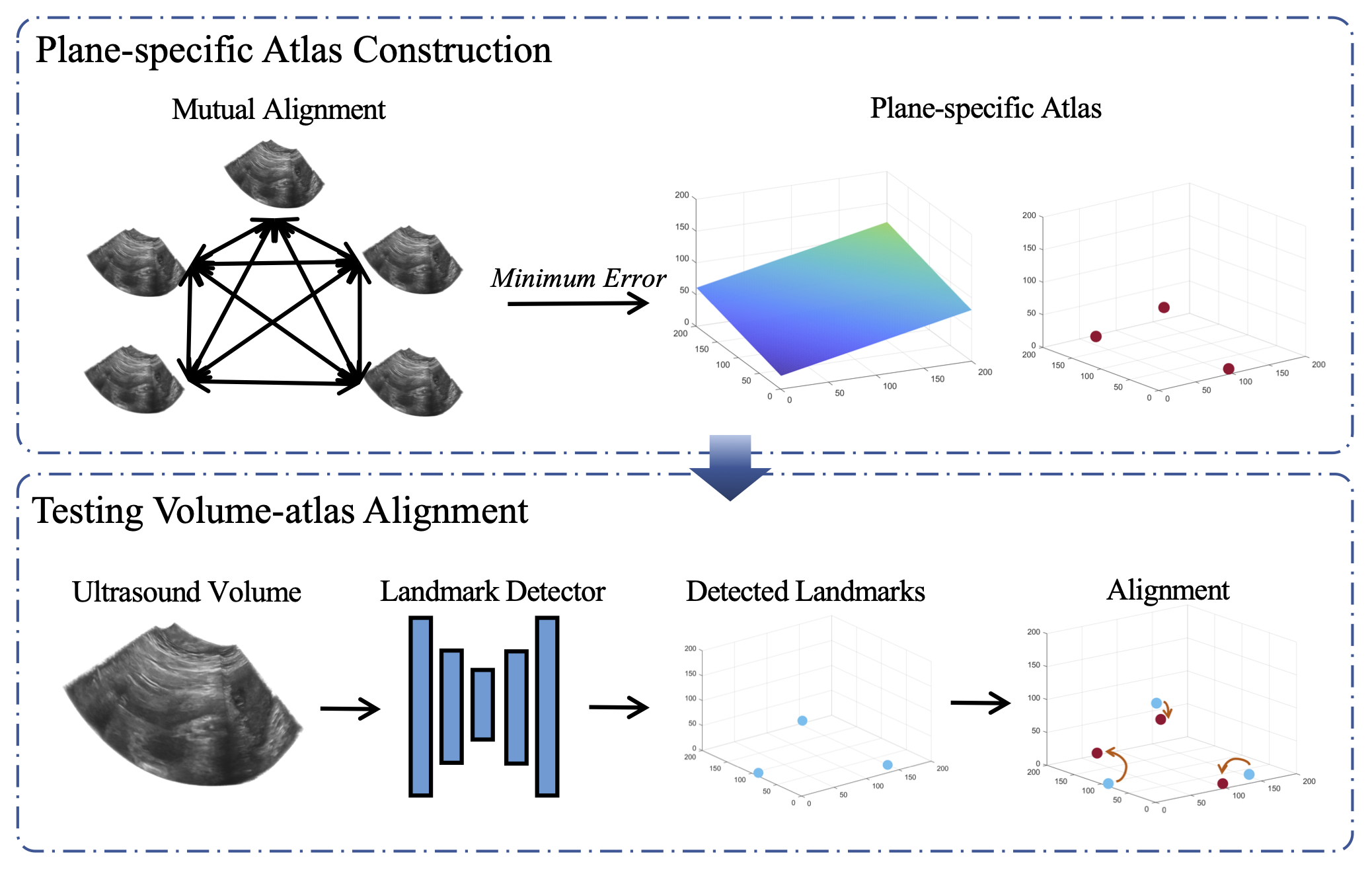}	
		\caption{Pipeline of our landmark-aware alignment module. We first select the plane-specific atlas from the training dataset. Then, a pre-trained detection network is employed to predict the landmarks. Finally, we obtain the bounded environment by aligning the testing volume with the atlas based on the landmarks. The bounded environment includes the aligned volume and the initial position of the standard plane.}
		\label{fig:warmup}
	\end{figure}
	
	Due to the low image quality, large data size and diverse fetal postures, it is very challenging to localize SPs in 3D US. Moreover, the random state initialization used in~\cite{Alansary10.1007/978-3-030-00928-1_32} often fails in localizing SPs because of the noisy 3D US environment. Therefore, a landmark-aware alignment module was proposed in our previous study~\cite{Dou10.1007/978-3-030-32254-0_33} as a dedicated warm-start of the searching process via anatomical prior knowledge. A more concrete processing pipeline is detailed in this section.
	
	This landmark-aware module aligns US volumes to the atlas space, thus reducing the diversity of fetal posture and US acquisition. As shown in Fig.~\ref{fig:warmup}, our proposed alignment module consists of two steps, namely plane-specific atlas construction and testing volume-atlas alignment. The details are described as follow.
	
	\begin{algorithm}[!hbtp]
		\caption{Atlas Selection}
		\small
		\label{alg:atlas_construction}
		\begin{algorithmic}[1]
			\FORALL {$i \in \left\{ 1,..., N \right\}$}
			\STATE {Refer to the US volume $V_{i}$ as the proxy atlas.}
			\FORALL {$j \in \left\{ 1,..., N \right\}$}
			\IF {$i == j$}
			\STATE {$continue$}
			\ELSE
			\STATE {$T_j^i\gets$ transformation matrix from $V_j$ to $V_i$.}
			\STATE {Perform rigid registration from $V_j$ to $V_i$.}
			\STATE {$\vec{n_{P}^{j} \gets}$ compute registered normal vector of plane $P$.}
			\STATE {$d^j_P \gets$} compute the distance from the origin to the plane $P$ in registered $V_{j}$.
			
			\STATE {$error_{i,j} = {\Theta}\footnotemark (\mathrm{T_{j}^{i}}\times \vec{n}_{P}^{j}, \vec{n}_{P}^{i}) +\parallel d_{P}^{j}-d_{P}^{i}\parallel_{1} $}
			\ENDIF
			\ENDFOR
			\STATE{$error_{i} = \frac{1}{N-1} \sum_{j}^{N-1}error_{i,j}$ }
			\ENDFOR
			\RETURN {The US volume with the minimum error.}
		\end{algorithmic}
		
	\end{algorithm}
	\footnotetext{$\Theta$ calculates the angle between plane normal vectors, referred to equation~\ref{eq:ang} in the Section~\ref{sec:evluation}.}

	$\bullet$ \textit{Plane-specific Atlas Construction:} In this study, the atlas is constructed to initialize the SP localization in the testing volume through landmark-based registration. Hence, the atlas selected from the training dataset must contain both reference landmarks for registration and SP parameters for plane initialization. As shown in Fig.~\ref{fig:warmup}, instead of selecting a common anatomical model for all SPs~\cite{Namburete2014DiagnosticPE,Lorenz10.1117/12.2292729}, we propose to select specific atlas for each SP to improve the localization accuracy. In order to ensure the initialization effectiveness, ideally, the specific SP of the selected atlas should be as close to the SPs of other training volumes as possible.  
	Algorithm~\ref{alg:atlas_construction} shows the determination of the plane-specific atlas volume from the training dataset based on minimum plane error (i.e. sum of the angle and distance between two planes). During the training stage, each volume is first taken as an initialized proxy atlas, then performing landmark-based rigid registration with the remaining volumes. According to the mean plane error measured between the linear-registered planes and ground truth for each proxy atlas, volume with the minimum error is chosen as the final atlas.
	
	$\bullet$ \textit{Testing Volume-atlas Alignment:} Our alignment module is based on landmark detection and matching. Unlike the direct regression, we convert the landmark detection as a heatmap regression task~\cite{huang2018vp} to avoid learning a highly abstract mapping function (i.e. feature representations to landmark coordinates). We trained a customized 3D U-net~\cite{iek20163DUL} with the L2-norm regression loss, denoted as:
	\begin{equation}
		\label{alignment label}
		\begin{split}
			{L} = \frac{1}{N}\sum_{n=1}^N\left(\mathcal{H}_i-\hat{\mathcal{H}}_i\right)^2
		\end{split}
	\end{equation}
	where $N=3$ denotes the number of landmarks, and $\mathcal{H}_i$, $\hat{\mathcal{H}}_i$ represent the $i$th predicted landmark heatmap and ground truth landmark heatmap, respectively. These ground truth heatmaps are created by placing a Gaussian kernel at the corresponding landmark location. During inference, we pass the test volumes to the landmark detector to get predicted landmark heatmaps. The coordinates with the highest value in the landmark heatmap are selected as the final prediction. We map the volume to the atlas space through the transform matrix calculated by the landmarks to create a bounded environment for the agent. Furthermore, we utilize the annotated target plane function of the atlas as the initial starting plane function for the agent.
	
	\subsection{Adaptive Dynamic Termination}	
	Compared with the current empirical termination strategy~\cite{Ghesu2019MultiScaleDR, Alansary10.1007/978-3-030-00928-1_32}, our previous work~\cite{Dou10.1007/978-3-030-32254-0_33} indicated that a learning-based termination strategy can improve the planning performance in deep RL. However, it requires the whole Q-value sequence obtained by maximum iterations to determine the final termination, which is inefficient. In this study, we update the active termination strategy into the adaptive dynamic termination, which is proposed in deep RL framework for the first time.

	Specifically, considering the sequential characteristics of the iterative interaction, as shown in Fig.~\ref{fig:framework}, we model the mapping between the Q-value sequence and optimal step with an additional RNN model. The Q-value is defined as $\mathbf{q}_{t}=\{q^{1}, q^{i}, ..., q^{8}\}$, consisting of 8 action candidates at the iteration $t$; and the Q-value sequence refers to a time-sequential matrix $\mathbf{Q}=[\mathbf{q}_1,\mathbf{q}_2, ..., \mathbf{q}_n]$, where $n$ denotes the index of iteration step. 
	Taking the Q-value sequence as input, the RNN model can learn the optimal termination step based on the highest Angle and Distance Improvement (ADI).\footnote{The definition of ADI refers to equation~\ref{eq:adi} in the Section~\ref{sec:evluation}.} During training, we randomly sampled the sub-sequences from the Q-value sequence as the training data and denoted the highest ADI during the sampling interval as the ground truth.

	Unlike the previous studies~\cite{Dou10.1007/978-3-030-32254-0_33, Alansary10.1007/978-3-030-00928-1_32}, we design a dynamic termination strategy to improve the inference efficiency of the reinforcement framework. Specifically, our RNN model performs one inference every two iterations based on the current zero-padding Q-value sequence, enabling an early stop at the iteration step having the first three repeated predictions.
	
	Our previous study~\cite{Dou10.1007/978-3-030-32254-0_33} used Mean Absolute Error (MAE) loss function to train the RNN in the termination module. However, it has constant gradient of back-propagation and lack of measuring the fine-grained error. This study replaces it with the Mean Square Error (MSE) loss function to relive this and target a more stable training procedure. Since ground truth, i.e. optimal termination step, is usually larger than 1 (e.g. 10$\sim$75), the conventional MSE loss function may struggle to converge in training due to the excessive gradient. We adopt the MSE loss function with a balance hyper-parameter, and defined as:

	\begin{equation}
		\label{eqa:RNN_loss}
		\begin{split}
			{L}_{MSE}\left({w}\right)=\left\|  {f}(x;{w})- \delta G_i \right\|_2^2
		\end{split}
	\end{equation}	
	where ${w}$ is the RNN parameters, $x$ is the input sequence of the RNN, $f\left(x; {w}\right)$ represents the RNN network, and $G$ denotes the optimal termination step. The balance hyper-parameter $\delta$ = 0.01 can normalize the value range of learned steps to [0, 0.75] approximately, thus simplifying the training process. The RNN model is trained using inference results obtained from training volumes. 
	% METHOD END
	
	% EXP AND RESULT START
	\section{Experiment configurations}
	\label{sec:experiment}
	\subsection{Implementation Details}
	\label{sec:implementation}
	We implemented our framework in PyTorch~\cite{Paszke2019PyTorchAI}, using a standard PC with an NVIDIA Titan XP GPU. We trained the whole framework through Adam optimizer~\cite{Kingma2015AdamAM} with a learning rate of 5e-5 and a batch size of 4 for 100 epochs, which cost about 4 days. We set the discount factor $\gamma$ in the loss function (Equation~\ref{eq:loss}) as 0.9. The size of the Replay Buffer was set as 15000. The target Q network copied the parameters of the Q network every 1500 iterations. The maximum number of iterations in one episode was set as 75 in fetal dataset and 30 in uterus dataset to reserve enough moving space for agent exploration. The initial $\epsilon$ for $\epsilon-greedy$ action selection strategy~\cite{Huang2017TemporalHT} was set as 0.6 at first and multiplied by 1.01 every 10000 iterations until 0.95 during training. The RNN variants, i.e. vanilla RNN and Long Short Term Memory (LSTM)~\cite{hochreiter1997long}, were trained for 100 epochs, using mini-batch Stochastic Gradient Descent (SGD)~\cite{bottou2007tradeoffs} optimizer with a learning rate of 1e-4 and batch size of 100, which costed about 45 mins. The number of hidden units was 64 and that of the RNN layers is 2. The starting plane function for training the framework was randomly initialized around the ground truth plane within an angle range of $25^\circ$ and distance range of $10~mm$ to ensure the agent can explore enough space within the US volume. For landmark detection, we trained the network using Adam optimizer with a batch size of 1 and a learning rate of 0.001 for 40 epochs.
	
	We chose the hyper-parameters based on the validation set and evaluate the performance of our method with several metrics on the held-out test sets. In specific, we trained the model for each hyper-parameters with different magnitudes and evaluate the performance on the validation dataset. We selected the value of hyper-parameters with the best validation performance as the default setup for the training phase. In this study, three high-impact hyper-parameters including the size of Replay Buffer, $\gamma$ and $\epsilon$ were searched.
	
	\subsection{Datasets}
	\label{sec:dataset}
	We validated the proposed framework using three distinct 3D US datasets, including fetal brain, fetal abdomen and uterus. Specifically, we aim to localize the three SPs: the transventricular (TV), the transthalamic (TT) and the transcerebellar (TC) SPs in the fetal brain, the fetal abdominal (AM) SP in the fetal abdomen, and the mid-sagittal (S), the transverse (T) and the coronal (C) SPs in uterus, respectively. We select three$/$four landmarks from each fetal$/$uterus US volume: the genu and the splenium of the corpus callosum, and the center of cerebellar vermis for fetal brain volumes; the umbilical vein entrance, the centrum, and the neck of the gallbladder for fetal abdomen volumes; two endometrial uterine horns, endometrial uterine bottom and uterine wall bottom for uterus volumes. We collected our dataset with 1635 prenatal 3D US volumes (433 fetal brains, 519 fetal abdomens and 683 uterus US volumes). Approved by the local Institutional Review Board, all volumes were anonymized and obtained by experts using a Mindray DC-9 ultrasound system with an integrated 3D probe. Average volume size of our dataset is $270 \times 207 \times 235$ in fetus and $261 \times 175 \times 277$ in uterus with a unified voxel size of $0.5\times0.5\times0.5{mm}^3$. Four sonographers with 5-year experience provided manual annotations of landmarks and SPs for all the volumes. All the annotation results were double-checked under strict quality control from a senior expert with 20-year experience. We randomly split our dataset for training, validating, testing of 313, 20, 100 in fetal brain, 389, 20, 110 in the fetal abdomen, and 519, 20, 144 in uterus.
	
	\subsection{Evaluation Criteria}
	\label{sec:evluation}
	In this study, we used three criteria to evaluate the localization accuracy of the predicted planes compared with the target plane. First, the angle and distance deviation between the two are estimated. Formally, we defined:
	\begin{equation}
		\label{eq:ang}
		\begin{split}
			Ang=\arccos \frac{{n}_{p}\cdot{n}_{g}}{\left|{n}_{p} \right| \left|{n}_{g} \right| }
		\end{split}
	\end{equation}
	\begin{equation}
		\label{eq:dis}
		\begin{split}
			Dis=\left|{d}_{p} - {d}_{g} \right|
		\end{split}
	\end{equation}
	where the ${n}_{p}, {n}_{g}$ represent the normal of the predict plane and target plane, the ${d}_{p}, {d}_{g} $ represent the distance from the volume origin to the predicted plane, and that to the ground truth plane. It is noted that the $Ang$ and $Dis$ are evaluated based on the plane sampling function, i.e., $cos(\alpha)x+cos(\beta)y +cos(\gamma)z=d$, with an effective voxel size of $0.5~mm^3/voxel$. Moreover, it is also important to examine whether these two planes are visually alike. Therefore, Peak Structural Similarity (SSIM)~\cite{Wang2004ImageQA} was leveraged to measure the image similarity of the planes.
	
	Besides, the ADI in iteration $t$ is defined as the sum of the cumulative changes of distance and angle from the start plane, which is as follows:
	\begin{equation}
			\label{eq:adi}
			ADI = ({Ang}_{t} - {Ang}_{0})+ ({Dis}_{t} - {Dis}_{0})
	\end{equation}
	
	\section{Results}
	In this section, we conducted  extensive experiments on the three dataset to validate the effectiveness and generalizability of our method. These experiments include performance comparison with state-of-the-art methods, effectiveness of the landmark-align module, effectiveness of the adaptive dynamic termination module, statistical significance test, clinical biometric evaluation, and qualitative evaluation. 
	
	\subsection{Comparison with state-of-the-art methods}
	To examine the effectiveness of our proposed method in standard plane localization, we conducted a comparison experiment with the classical learning-based regression method, denoted as \textit{Regression}, the current state-of-the-art Automatic View Planning method~\cite{Alansary10.1007/978-3-030-00928-1_32}, denoted as \textit{AVP}, and our previous method~\cite{Dou10.1007/978-3-030-32254-0_33}, denoted as \textit{RL-US}. To achieve a fair comparison, we used the default plane initialization strategy of the \textit{Regression} and \textit{AVP}, and re-trained all the two compared models using the public implementations. We also adjusted the training parameters to obtain the best localization results. As shown in Table~\ref{tab:exist_method_fetal} and~\ref{tab:exist_method_uterus}, it can be observed that our method achieves the highest accuracy compared with the alternatives in almost all of the metrics. This indicates the superior ability of our method in standard plane localization tasks.
	
	\begin{table*}[!htbp]
		\scriptsize
		\centering
		\caption{comparison results of our proposed method and other existing methods in \textbf{fetal US} (mean$\pm$std, best results are highlighted in bold).}
		\label{tab:exist_method_fetal}
		\setlength{\tabcolsep}{0.75mm}{
			\begin{tabular}{c c c c c c c c c c c c c}
				\hline
				\hline
				\multirow{2}{*}{\bf{Method}} & \multicolumn{3}{c}{\bf{TC}}  & \multicolumn{3}{c}{\bf{TV}}  & \multicolumn{3}{c}{\bf{TT}}  & \multicolumn{3}{c}{\bf{AM}}\\
				\cline{2-13}
				&Ang($^{\circ}$) &Dis(mm) &SSIM	
				&Ang($^{\circ}$) &Dis(mm) &SSIM
				&Ang($^{\circ}$) &Dis(mm) &SSIM
				&Ang($^{\circ}$) &Dis(mm) &SSIM \\
				\hline
				Regression
				&12.44$\pm$7.78 &\textbf{2.18$\pm$2.12} &0.634$\pm$0.157
				&13.62$\pm$4.98 &5.52$\pm$4.02 &0.636$\pm$0.136
				&13.87$\pm$11.77 &2.81$\pm$2.16 &0.760$\pm$0.141
				&17.48$\pm$12.45 &6.24$\pm$4.77 &0.758$\pm$0.081\\
				AVP~\cite{Alansary10.1007/978-3-030-00928-1_32}
				&48.42$\pm$12.45 &10.55$\pm$7.46 &0.580$\pm$0.047
				&48.31$\pm$18.25 &14.64$\pm$10.20 &0.586$\pm$0.054
				&57.35$\pm$12.31 &9.92$\pm$6.29 &0.554$\pm$0.088
				&46.52$\pm$13.54 &7.71$\pm$7.01 &0.649$\pm$0.071 \\
				\hline
				RL-US~\cite{Dou10.1007/978-3-030-32254-0_33}
				&10.54$\pm$9.45 &2.55$\pm$2.45 &0.639$\pm$0.131
				&10.40$\pm$8.46 &2.65$\pm$1.62 &0.655$\pm$0.131
				&\textbf{10.37$\pm$8.08} &3.46$\pm$2.89 &0.769$\pm$0.087
				&14.84$\pm$8.22 &2.42$\pm$1.96 &0.784-0.080  \\
				Ours
				&\textbf{10.26$\pm$7.25} &2.52$\pm$2.13 &\textbf{0.640$\pm$0.144}
				&\textbf{10.39$\pm$4.03} &\textbf{2.48$\pm$1.27} &\textbf{0.659$\pm$0.135}
				&10.48$\pm$5.80 &\textbf{2.02$\pm$1.33} &\textbf{0.783$\pm$0.060}
				&\textbf{14.57$\pm$8.50} &\textbf{2.00$\pm$1.64} &\textbf{0.790$\pm$0.074} \\
				\hline
				\hline
		\end{tabular}}
	\end{table*}
	
	\begin{table*}[!htbp]
		\scriptsize
		\centering
		\caption{comparison results of our proposed method and other existing methods in \textbf{Uterus US} (mean$\pm$std, best results are highlighted in bold).}
		\label{tab:exist_method_uterus}
		\setlength{\tabcolsep}{0.75mm}{
			\begin{tabular}{c c c c c c c c c c}
				\hline
				\hline
				\multirow{2}{*}{\bf{Method}} & \multicolumn{3}{c}{\bf{S}}  & \multicolumn{3}{c}{\bf{T}}  & \multicolumn{3}{c}{\bf{C}} \\
				\cline{2-10}
				&Ang($^{\circ}$) &Dis(mm) &SSIM	
				&Ang($^{\circ}$) &Dis(mm) &SSIM
				&Ang($^{\circ}$) &Dis(mm) &SSIM\\
				\hline
				Regression
				&10.27$\pm$11.40 &3.34$\pm$1.30 &0.835$\pm$0.092
				&10.57$\pm$8.84 &\textbf{3.03$\pm$2.71} &0.693$\pm$0.112
				&9.07$\pm$7.85 &2.23$\pm$1.11 &0.605$\pm$0.094\\
				AVP~\cite{Alansary10.1007/978-3-030-00928-1_32}
				&44.12$\pm$14.00 &15.27$\pm$9.87 &0.736$\pm$0.066
				&66.71$\pm$17.31 &16.48$\pm$13.02 &0.575$\pm$0.073
				&53.21$\pm$13.21 &13.41$\pm$14.47 &0.522$\pm$0.064 \\
				\hline	
				RL-US~\cite{Dou10.1007/978-3-030-32254-0_33}
				&9.89$\pm$9.73 &2.62$\pm$2.99 &0.886$\pm$0.061
				&9.59$\pm$8.24 &3.08$\pm$2.40 &0.773$\pm$0.094
				&\textbf{7.50$\pm$6.69} &1.56$\pm$1.48 &0.697$\pm$0.096\\
				Ours
				&\textbf{9.71$\pm$9.69} &\textbf{2.61$\pm$3.02} &\textbf{0.888$\pm$0.062}
				&\textbf{9.58$\pm$8.08} &3.09$\pm$2.38 &\textbf{0.773$\pm$0.092}
				&7.54$\pm$6.64 &\textbf{1.49$\pm$1.47} &\textbf{0.699$\pm$0.098}\\
				\hline
				\hline
		\end{tabular}}
	\end{table*}
	
	\subsection{Impacts of the landmark-align module}
	\begin{figure}[!htbp]
	\centering
	\includegraphics[width=0.95\linewidth]{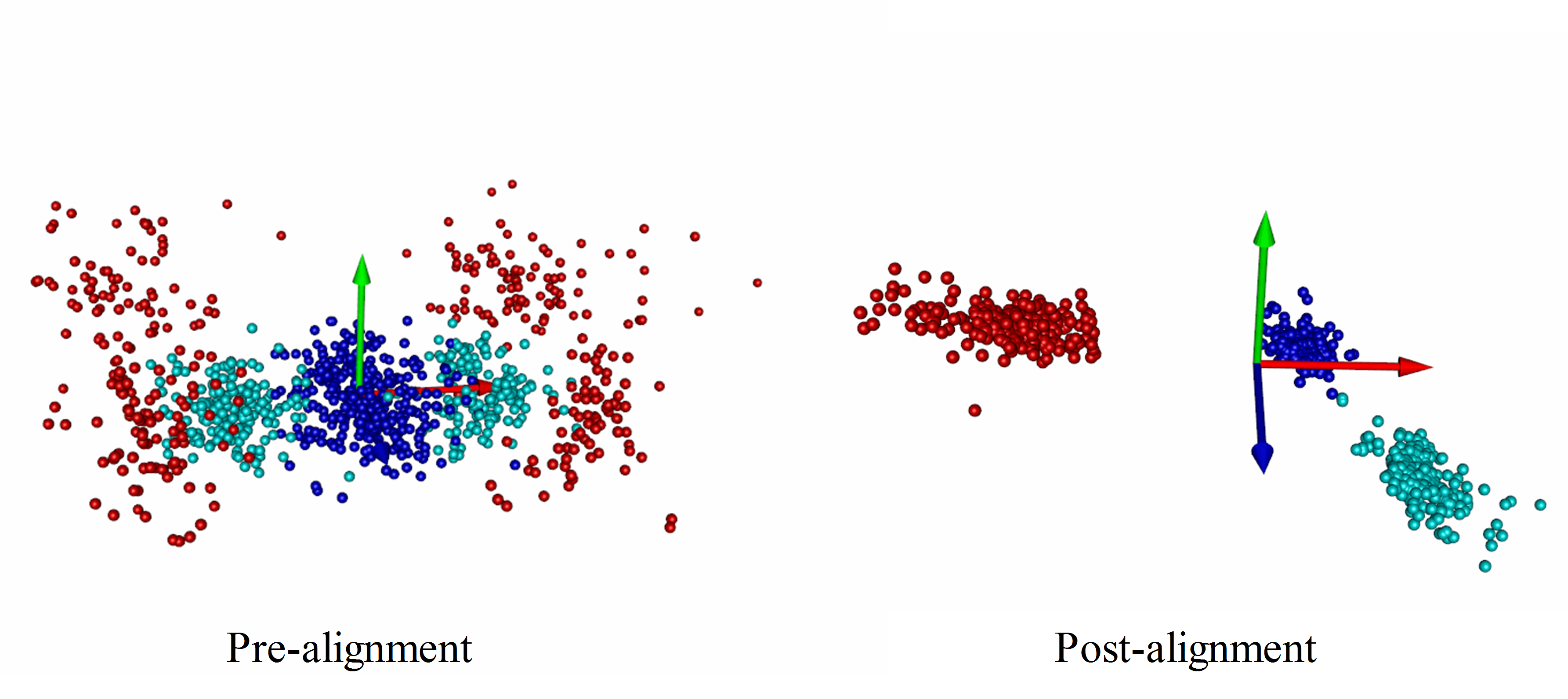}	
	\caption{The 3D visualization of fetal brain landmarks distribution of pre- and post-alignment. Different color points represent different category landmarks.}
	\label{fig:landmark_acc}
	\end{figure}
	To verify the impact of the landmark-aware alignment module of the proposed approach, we compared the performance of the proposed framework with and without this module. In the \textit{Pre-Regist} method, we set the agent with random starting plane function like~\cite{Alansary10.1007/978-3-030-00928-1_32} and choose the lowest Q-value~\cite{Alansary10.1007/978-3-030-00928-1_32} as the termination step. The \textit{Regist} method represents the framework equipped with the alignment module but without agent searching. The \textit{Post-Regist} method denotes the searching result of the agent with a warm-up initialization with the alignment module. We also chose the lowest Q-value termination strategy to implement the \textit{Post-Regist} for a fair comparison. As shown in the Table~\ref{tab:warm_up_fetal} and~\ref{tab:warm_up_uterus}, the accuracy of the \textit{Pre-Regist} method is significantly lower than that of the \textit{Regist} and the \textit{Post-Regist} method. This proves that the landmark-aware alignment module can improve the plane detection accuracy consistently and substantially. Figure~\ref{fig:landmark_acc} provides visualization of the 3D spatial distribution of the fetal brain landmarks pre-/post-alignment. It can be observed that all the landmarks are mapped to a similar spatial position, which indicates that all the fetal postures are roughly aligned.

	\begin{table*}[!htbp]
		\scriptsize
		\centering
		\caption{comparison results of the ablation study for analysis of warm start in \textbf{fetal US} (mean$\pm$std, best results are highlighted in bold).}
		\label{tab:warm_up_fetal}
		\setlength{\tabcolsep}{0.75mm}{
			\begin{tabular}{c c c c c c c c c c c c c}
				\hline
				\hline
				\multirow{2}{*}{\bf{Method}} & \multicolumn{3}{c}{\bf{TC}}  & \multicolumn{3}{c}{\bf{TV}}  & \multicolumn{3}{c}{\bf{TT}}  & \multicolumn{3}{c}{\bf{AM}}\\
				\cline{2-13}
				&Ang($^{\circ}$) &Dis(mm) &SSIM	
				&Ang($^{\circ}$) &Dis(mm) &SSIM
				&Ang($^{\circ}$) &Dis(mm) &SSIM
				&Ang($^{\circ}$) &Dis(mm) &SSIM\\
				\hline
				Pre-Regist
				&48.42$\pm$12.45 &10.55$\pm$7.46 &0.580$\pm$0.047
				&48.31$\pm$18.25 &14.64$\pm$10.20 &0.586$\pm$0.054
				&57.35$\pm$12.31 &9.92$\pm$6.29 &0.554$\pm$0.088
				&46.52$\pm$13.54 &7.71$\pm$7.01 &0.649$\pm$0.071 \\
				
				Regist
				&14.28$\pm$7.62 &3.48$\pm$2.41 &0.609$\pm$0.129
				&13.96$\pm$4.33 &\textbf{2.39$\pm$1.34} &0.642$\pm$0.120
				&14.36$\pm$13.41 &\textbf{2.11$\pm$1.41} &0.767$\pm$0.079
				&17.31$\pm$12.04 &2.57$\pm$2.34 &0.773$\pm$0.080 \\
				\hline
				Post-Regist
				&\textbf{10.98$\pm$9.86} &\textbf{2.88$\pm$2.46} &\textbf{0.636$\pm$0.144}
				&\textbf{11.30$\pm$10.80} &2.66$\pm$1.69 &\textbf{0.649$\pm$0.128}
				&\textbf{12.28$\pm$8.77} &2.62$\pm$2.50 &\textbf{0.769$\pm$0.071}
				&\textbf{16.05$\pm$8.93} &\textbf{2.24$\pm$2.10} &\textbf{0.776$\pm$0.079}  \\
				\hline
				\hline
		\end{tabular}}
	\end{table*}
	
		\begin{table*}[!htbp]
			\scriptsize
			\centering
			\caption{comparison results of the ablation study for analysis of warm start in \textbf{uterus US} (mean$\pm$std, best results are highlighted in bold).}
			\label{tab:warm_up_uterus}
			\setlength{\tabcolsep}{0.75mm}{
				\begin{tabular}{c c c c c c c c c c}
					\hline
					\hline
					\multirow{2}{*}{\bf{Method}} & \multicolumn{3}{c}{\bf{S}}  & \multicolumn{3}{c}{\bf{T}}  & \multicolumn{3}{c}{\bf{C}} \\
					\cline{2-10}
					&Ang($^{\circ}$) &Dis(mm) &SSIM	
					&Ang($^{\circ}$) &Dis(mm) &SSIM
					&Ang($^{\circ}$) &Dis(mm) &SSIM\\
					\hline
					Pre-Regist
					&44.12$\pm$14.00 &15.27$\pm$9.87 &0.736$\pm$0.066
					&66.71$\pm$17.31 &16.48$\pm$13.02 &0.575$\pm$0.073
					&53.21$\pm$13.21 &13.41$\pm$14.47 &0.522$\pm$0.064 \\
					Regist
					&11.55$\pm$9.99 &2.86$\pm$3.39 &0.814$\pm$0.082
					&9.96$\pm$7.18 &3.29$\pm$2.60 &0.671$\pm$0.103
					&8.50$\pm$6.73 &\textbf{1.54$\pm$1.60} &0.545$\pm$0.127 \\
					\hline
					Post-Regist
					&\textbf{9.85$\pm$9.74} &\textbf{2.56$\pm$3.03} &\textbf{0.884$\pm$0.066}
					&\textbf{9.72$\pm$8.08} &\textbf{3.10$\pm$2.55} &\textbf{0.770$\pm$0.105}
					&\textbf{7.48$\pm$6.43} &1.70$\pm$1.56 &\textbf{0.686$\pm$0.093}\\
					\hline
					\hline
			\end{tabular}}
	\end{table*}
	
	\subsection{Analysis of adaptive dynamic termination}
	
	\begin{table*}[!htbp]
		\scriptsize
		\centering
		\caption{comparison results of the ablation study for analysis of termination strategy in \textbf{fetal US} (mean$\pm$std, best results are highlighted in bold).}
		\label{tab:termination_fetal}
		\setlength{\tabcolsep}{0.60mm}{
			\begin{tabular}{c c c c c c c c c c c c c}
				\hline
				\hline
				\multirow{2}{*}{\bf{Method}} & \multicolumn{3}{c}{\bf{TC}}  & \multicolumn{3}{c}{\bf{TV}}  & \multicolumn{3}{c}{\bf{TT}}  & \multicolumn{3}{c}{\bf{AM}}\\
				\cline{2-13}
				&Ang($^{\circ}$) &Dis(mm) &SSIM	
				&Ang($^{\circ}$) &Dis(mm) &SSIM
				&Ang($^{\circ}$) &Dis(mm) &SSIM
				&Ang($^{\circ}$) &Dis(mm) &SSIM\\
				\hline
				Max-Step
				&12.73$\pm$11.33 &3.74$\pm$3.33 &0.619$\pm$0.135
				&12.17$\pm$11.54 &2.80$\pm$1.85 &0.645$\pm$0.127
				&17.29$\pm$14.56 &4.76$\pm$5.17 &0.727$\pm$0.091
				&18.97$\pm$11.93 &2.90$\pm$3.10 &0.755$\pm$0.095  \\
				
				Low Q-Value~\cite{Alansary10.1007/978-3-030-00928-1_32}
				&10.98$\pm$9.86 &2.88$\pm$2.46 &0.636$\pm$0.144
				&11.30$\pm$10.80 &2.66$\pm$1.69 &0.649$\pm$0.128
				&12.28$\pm$8.77 &2.62$\pm$2.50 &0.769$\pm$0.071
				&16.05$\pm$8.93 &2.24$\pm$2.10 &0.776$\pm$0.079  \\
				
				AT-LSTM~\cite{Dou10.1007/978-3-030-32254-0_33}
				&10.54$\pm$9.45 &2.55$\pm$2.45 &0.639$\pm$0.131
				&10.40$\pm$8.46 &2.65$\pm$1.62 &0.655$\pm$0.131
				&\textbf{10.37$\pm$8.08} &3.46$\pm$2.89 &0.769$\pm$0.087
				&14.84$\pm$8.22 &2.42$\pm$1.96 &0.784-0.080  \\

				\hline
				ADT-MLP
				&10.39$\pm$7.33 &2.55$\pm$2.15 &0.640$\pm$0.145
				&10.97$\pm$4.56 &2.57$\pm$1.49 &0.653$\pm$0.133
				&11.90$\pm$7.50 &3.20$\pm$3.43 &0.764$\pm$0.074
				&15.30$\pm$8.27 &2.34$\pm$2.18 &0.779$\pm$0.081\\
				
				ADT-RNN
				&10.63$\pm$7.24 &2.66$\pm$2.19 &0.640$\pm$0.142
				&10.90$\pm$4.46 &2.55$\pm$1.47 &0.655$\pm$0.134
				&11.45$\pm$7.18 &2.78$\pm$3.12 &0.774$\pm$0.068
				&15.05$\pm$0.77 &2.26$\pm$2.06 &0.781$\pm$0.077\\
				
				ADT-LSTM
				&10.49$\pm$7.33 &2.56$\pm$2.14 &0.639$\pm$0.144
				&10.60$\pm$4.30 &2.49$\pm$1.44 &0.657$\pm$0.136
				&10.84$\pm$6.23 &2.64$\pm$2.88 &0.775$\pm$0.066
				&14.92$\pm$8.08 &2.28$\pm$2.15 &0.784$\pm$0.076\\
				ADT-LSTM$^*$
				&\textbf{10.26$\pm$7.25} &\textbf{2.52$\pm$2.13} &\textbf{0.640$\pm$0.144}
				&\textbf{10.39$\pm$4.03} &\textbf{2.48$\pm$1.27} &\textbf{0.659$\pm$0.135}
				&10.48$\pm$5.80 &\textbf{2.02$\pm$1.33} &\textbf{0.783$\pm$0.060}
				&\textbf{14.57$\pm$8.50} &\textbf{2.00$\pm$1.64} &\textbf{0.790$\pm$0.074} \\
				\hline
				\hline
		\end{tabular}}
	\end{table*}

	\begin{table*}[!htbp]
		\scriptsize
		\centering
		\caption{comparison results of the ablation study for analysis of termination strategy in UTERUS US (mean$\pm$std, best results are highlighted in bold).}
		\label{tab:termination_uterus}
		\setlength{\tabcolsep}{0.60mm}{
			\begin{tabular}{c c c c c c c c c c}
				\hline
				\hline
				\multirow{2}{*}{\bf{Method}} & \multicolumn{3}{c}{\bf{S}}  & \multicolumn{3}{c}{\bf{T}}  & \multicolumn{3}{c}{\bf{C}} \\
				\cline{2-10}
				&Ang($^{\circ}$) &Dis(mm) &SSIM	
				&Ang($^{\circ}$) &Dis(mm) &SSIM
				&Ang($^{\circ}$) &Dis(mm) &SSIM\\
				\hline
				Max-Step
				&9.82$\pm$9.12 &2.70$\pm$3.21 &0.877$\pm$0.070
				&9.83$\pm$8.72 &3.11$\pm$2.66 &0.770$\pm$0.101
				&8.14$\pm$6.84 &1.99$\pm$1.67 &0.670$\pm$0.090\\
				Low Q-Value~\cite{Alansary10.1007/978-3-030-00928-1_32}
				&9.85$\pm$9.74 &\textbf{2.56$\pm$3.13} &0.884$\pm$0.066
				&9.72$\pm$8.08 &3.10$\pm$2.55 &0.770$\pm$0.105
				&\textbf{7.48$\pm$6.43} &1.70$\pm$1.56 &0.686$\pm$0.093\\
				
				AT-LSTM~\cite{Dou10.1007/978-3-030-32254-0_33}
				&9.89$\pm$9.73 &2.62$\pm$2.99 &0.886$\pm$0.061
				&9.59$\pm$8.24 &3.08$\pm$2.40 &0.773$\pm$0.094
				&7.50$\pm$6.69 &1.56$\pm$1.48 &0.697$\pm$0.096\\
				
				\hline
				ADT-MLP
				&9.89$\pm$9.76 &2.63$\pm$2.99 &0.885$\pm$0.060
				&9.70$\pm$8.43 &\textbf{3.07$\pm$2.47} &0.772$\pm$0.097
				&7.69$\pm$6.65 &1.59$\pm$1.50 &0.698$\pm$0.094\\
				
				ADT-RNN
				&9.85$\pm$9.76 &2.62$\pm$3.00 &0.886$\pm$0.061
				&9.66$\pm$8.40 &3.08$\pm$2.44 &0.773$\pm$0.097
				&7.65$\pm$6.63 &1.52$\pm$1.48 &0.697$\pm$0.093\\
				
				ADT-LSTM
				&10.03$\pm$9.73 &2.63$\pm$2.97 &0.885$\pm$0.059
				&9.60$\pm$8.10 &3.09$\pm$2.39 &0.773$\pm$0.093
				&7.54$\pm$6.60 &1.53$\pm$1.47 &0.698$\pm$0.096\\
				
				ADT-LSTM$^*$
				%&9.96$\pm$9.75 &2.63$\pm$2.98 &0.885$\pm$0.060
				&\textbf{9.71$\pm$9.69} &2.61$\pm$3.02 &\textbf{0.888$\pm$0.062}
				&\textbf{9.58$\pm$8.08} &3.09$\pm$2.38 &\textbf{0.773$\pm$0.092}
				&7.54$\pm$6.64 &\textbf{1.49$\pm$1.47} &\textbf{0.699$\pm$0.098}\\
				
				\hline
				\hline
		\end{tabular}}
	\end{table*}

	\begin{table}[htbp]
		\centering
		\caption{Average Termination Step of the Adaptive dynamic termination and active termination}
		\label{tab:termination_step}
		\begin{tabular}{c c c}
			\hline
			\hline
			\multirow{2}{*}{Plane} &\multicolumn{2}{c}{Termination Step} \\
			\cline{2-3} 
			&AT &ADT \\
			\hline
			TC
			&75	&39.0\\
			TV
			&75 &24.8\\
			TT
			&75 &18.8\\
			AM
			&75 &20.3\\
			\hline
			S
			&30 &18.0\\
			T
			&30 &27.8\\
			C
			&30 &7.5\\
			\hline
			\hline
		\end{tabular}
	\end{table}

	To demonstrate the impact of the proposed adaptive dynamic termination (\textit{ADT}) strategy, we performed comparison experiments with existing popular strategies such as the termination with max iterations (\textit{Max-Step}), the lowest Q Value (\textit{Low Q-Value}~\cite{Alansary10.1007/978-3-030-00928-1_32}), and the active termination~\cite{Dou10.1007/978-3-030-32254-0_33} with LSTM (\textit{AT-LSTM}). We also compared with our proposed \textit{ADT} with different backbone network including Multi-layer Perceptron (\textit{ADT-MLP}), vanilla RNN (\textit{ADT-RNN}) and LSTM (\textit{ADT-LSTM}). 
	The superscript $*$ represents the model was trained with the normalized MSE loss function ($L_{MSE}$, Eq.~\ref{eqa:RNN_loss})
	As shown in Table~\ref{tab:termination_fetal} and~\ref{tab:termination_uterus}, equipped with the adaptive dynamic termination strategy, the agent was able to avoid being trapped into an inferior local minimum and achieved better performance. Furthermore, from Table~\ref{tab:termination_step}, we can observe that our proposed dynamic termination can save approximately $67\%$ inference time at most, thus improving the efficiency of the reinforcement framework.
	
	\begin{figure}[!htbp]
		\centering
		\includegraphics[width=0.8\linewidth]{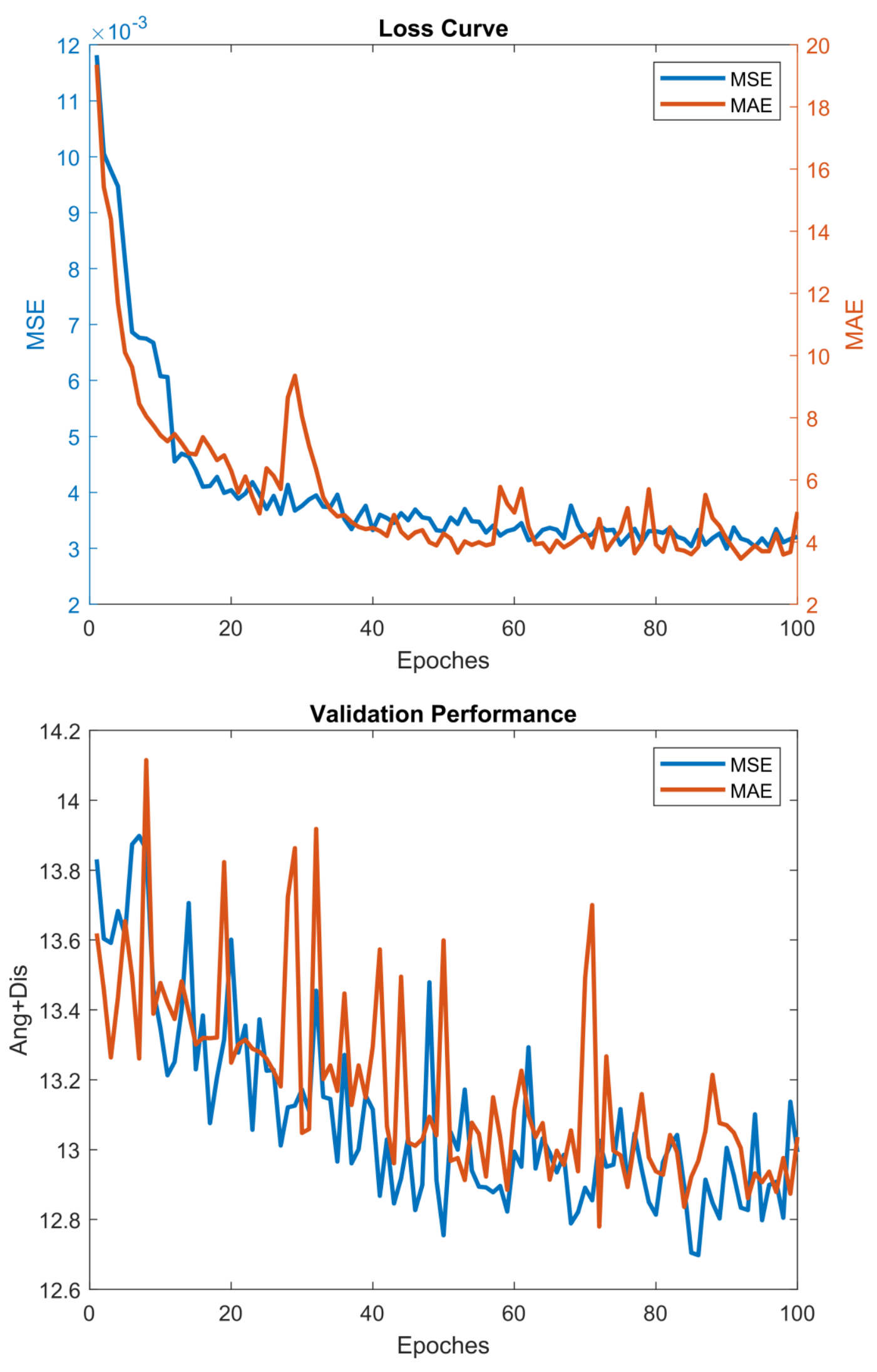}	
		\caption{The training curves and validation performance of the termination module trained with MAE loss and modified M-MSE loss. The top is the loss curves and the bottom is the curves of sum of angle and distance in validation. We performed the training and validation sequentially every epoch, where we obtained the loss and validation performance curves.}
		\label{fig:l2vsl1}
	\end{figure}
	
	Figure~\ref{fig:l2vsl1} displays the training curves and validation performance of the same model trained with the MAE loss~\cite{Dou10.1007/978-3-030-32254-0_33} and the normalized MSE loss, respectively. It shows that the MSE loss can facilitate the model to obtain a more stable convergence and lower $Ang+Dis$ in validation as comparison to the MAE loss. This indicates the effectiveness of the $L_{MSE}$ in simplifying the training of the termination module.
	
	As shown in Table~\ref{tab:lstm_abltation}, we performed the ablation study of the number of layers and hidden units of the LSTM in fetal brain dataset. We can observe that the LSTM with 2 layers and 64 hidden units outperforms those with other settings. 

	\begin{table*}[!htbp]
		\scriptsize
		\centering
		\caption{Ablation Study for the number of layers and hidden uints of the LSTM in fetal brain dataset.}
		\label{tab:lstm_abltation}
		\setlength{\tabcolsep}{0.60mm}{
			\begin{tabular}{c c c c c c c c c c c}
				\hline
				\hline
				\multirow{2}{*}{\bf{Num of layers}} &\multirow{2}{*}{\bf{Num of units}} & \multicolumn{3}{c}{\bf{TC}}  & \multicolumn{3}{c}{\bf{TV}}  & \multicolumn{3}{c}{\bf{TT}} \\
				\cline{3-11}
				& &Ang($^{\circ}$) &Dis(mm) &SSIM	
				&Ang($^{\circ}$) &Dis(mm) &SSIM
				&Ang($^{\circ}$) &Dis(mm) &SSIM\\
				\hline
				1 & 64
				&10.37$\pm$7.14 &2.71$\pm$2.11 &\textbf{0.641$\pm$0.145}
				&10.67$\pm$3.72 &\textbf{2.44$\pm$1.48} &0.657$\pm$0.131
				&10.90$\pm$6.17 &2.70$\pm$2.98 &0.774$\pm$0.066\\
				2 & 64
				&\textbf{10.36$\pm$7.26} &2.56$\pm$2.14 &0.638$\pm$0.143
				&\textbf{10.39$\pm$4.03} &2.48$\pm$1.27 &\textbf{0.659$\pm$0.135}
				&\textbf{10.72$\pm$6.18} &2.59$\pm$2.77 &0.777$\pm$0.064\\
				4 & 64
				&10.41$\pm$7.29 &\textbf{2.52$\pm$2.11} &0.637$\pm$0.141
				&10.49$\pm$4.01 &2.50$\pm$1.28 &0.658$\pm$0.131
				&10.78$\pm$6.05 &2.61$\pm$2.92 &\textbf{0.778$\pm$0.064}\\
				\hline
				2 & 16
				&10.41$\pm$7.28 &2.55$\pm$2.14 &0.638$\pm$0.142
				&10.57$\pm$3.79 &2.49$\pm$1.44 &0.658$\pm$0.134
				&11.32$\pm$6.89 &\textbf{2.55$\pm$2.32} &0.777$\pm$0.067\\
				2 & 256
				&10.50$\pm$7.33 &2.56$\pm$2.13 &0.635$\pm$0.140
				&10.51$\pm$3.84 &2.49$\pm$1.43 &0.657$\pm$0.134
				&10.88$\pm$5.97 &2.61$\pm$2.88 &0.777$\pm$0.064\\
				
				\hline
				\hline
		\end{tabular}}
	\end{table*}

	\subsection{Significant Difference Analysis}
	To investigate if the difference between methods were statistically significant, we performed paired \textit{t}-tests between the results of our methods and \textit{Regression}, \textit{AVP}~\cite{Alansary10.1007/978-3-030-00928-1_32}, \textit{Registration}. These tests were conducted for all of the performance metrics including \textit{Angle}, \textit{Distance} and \textit{SSIM}. We set the significance level as 0.05.  The results are shown in the Table~\ref{tab:t-test-fetal} and~\ref{tab:t-test-uterus}. The results of the comparisons and tests in Tables I-IV and~\ref{tab:t-test-fetal}-\ref{tab:t-test-uterus} indicate that our method performed best among the state-of-art methods (\textit{Regression}, \textit{AVP}~\cite{Alansary10.1007/978-3-030-00928-1_32}) and \textit{Registration}. Although our method outperforms the \textit{AT-LSTM}~\cite{Dou10.1007/978-3-030-32254-0_33} without significant difference, our method could save at most 67 $\%$ inference time as shown in Table~\ref{tab:termination_step}. 
	
	\begin{table*}[!htbp]
		\scriptsize
		\centering
		\caption{\textit{p}-values of pairwise \textit{t}-tests between the results of each method and our method for the three performance metrics in the fetal dataset. The Bolded results represent significant difference}
		\label{tab:t-test-fetal}
		\setlength{\tabcolsep}{0.70mm}{
			\begin{tabular}{c c c c c c c c c c c c c}
				\hline
				\hline
				\multirow{2}{*}{\bf{Metric}} & \multicolumn{3}{c}{\bf{TC}}  & \multicolumn{3}{c}{\bf{TV}}  & \multicolumn{3}{c}{\bf{TT}}  & \multicolumn{3}{c}{\bf{AM}}\\
				\cline{2-13}
				&Ang($^{\circ}$) &Dis(mm) &SSIM	
				&Ang($^{\circ}$) &Dis(mm) &SSIM
				&Ang($^{\circ}$) &Dis(mm) &SSIM
				&Ang($^{\circ}$) &Dis(mm) &SSIM \\
				\hline
				Ours vs. Regression
				&$\bf10^{-4}$ &\textbf{0.003} &0.138
				&\textbf{0.003} &$\bf10^{-15}$ &0.282
				&$\bf10^{-4}$ &\textbf{0.008} &0.301
				&\textbf{0.006} &$\bf10^{-25}$ &0.161 \\
				Ours vs. AVP~\cite{Alansary10.1007/978-3-030-00928-1_32}
				&$\bf10^{-56}$ &$\bf10^{-45}$ &\textbf{0.003} 
				&$\bf10^{-57}$ &$\bf10^{-56}$ &\textbf{0.001} 
				&$\bf10^{-64}$ &$\bf10^{-44}$ &$\bf10^{-18}$ 
				&$\bf10^{-49}$ &$\bf10^{-32}$ &$\bf10^{-10}$  \\
				Ours vs. Registration
				&$\bf10^{-4}$ &\textbf{0.003}  &0.162
				&\textbf{0.001} &0.454 &0.399
				&$\bf10^{-4}$ &\textbf{0.049} &0.329
				&\textbf{0.005} &\textbf{0.048} &0.399  \\
				\hline
				\hline
		\end{tabular}}
	\end{table*}

		\begin{table*}[!htbp]
			\scriptsize
			\centering
			\caption{\textit{p}-values of pairwise \textit{t}-tests between the results of each method and our method for the three performance metrics in the uterus dataset. The Bolded results represent significant difference}
			\label{tab:t-test-uterus}
			\setlength{\tabcolsep}{0.75mm}{
				\begin{tabular}{c c c c c c c c c c}
					\hline
					\hline
					\multirow{2}{*}{\bf{Metric}} & \multicolumn{3}{c}{\bf{S}}  & \multicolumn{3}{c}{\bf{T}}  & \multicolumn{3}{c}{\bf{C}}  \\
					\cline{2-10}
					&Ang($^{\circ}$) &Dis(mm) &SSIM	
					&Ang($^{\circ}$) &Dis(mm) &SSIM
					&Ang($^{\circ}$) &Dis(mm) &SSIM\\
					\hline
					Ours vs. Regression
					&0.603 &\textbf{0.009} &\textbf{0.005} 
					&0.276 &0.848 &$\bf10^{-5}$
					&0.130 &\textbf{0.025} &$\bf10^{-6}$  \\
					Ours vs. AVP~\cite{Alansary10.1007/978-3-030-00928-1_32}
					&$\bf10^{-49}$ &$\bf10^{-63}$ &$\bf10^{-11}$
					&$\bf10^{-71}$ &$\bf10^{-63}$ &$\bf10^{-15}$
					&$\bf10^{-61}$ &$\bf10^{-58}$ &$\bf10^{-14}$ \\
					Ours vs. Registration
					&\textbf{0.047} &0.359 &$\bf10^{-4}$
					&0.720 &0.495 &$\bf10^{-6}$
					&0.329 &0.877 &$\bf10^{-10}$\\
					\hline
					\hline
			\end{tabular}}
	\end{table*}

	\subsection{Clinical biometric evaluation from SP}

	\begin{table}[htbp]
		\centering
		\caption{Segmentation performance and the Quantitative analysis of the clinical assessment.}
		\label{tab:seg}
		\begin{tabular}{c c c c c}
			\hline
			\hline
			Plane &Dice &A-Error(mm) & R-Error($\%$) &p-value\\
			\hline
			TT
			&0.971$\pm$0.009 &1.125$\pm$1.431 &2.05$\pm$1.56 &0.22\\
			AM
			&0.941$\pm$0.047 &3.608$\pm$3.462 &3.25$\pm$4.44 &0.13\\
			\hline
			\hline
		\end{tabular}
	\end{table}
	
	In this section, we further explore whether the detected planes can provide accurate biometrics that are consistent with the ones obtained in manually acquired planes, which are more of clinical concerns. 
	To obtain those on the predicted planes (TT and AM), a pre-trained DeepLabv3+~\cite{Chen2018EncoderDecoderWA} was used to perform segmentation of fetal head and abdomen. Then two smallest ellipses enclosing the segmentation map in predicted plane and the annotated ground truth in target plane are generated for the fetal head or abdominal circumference. We used three metrics to evaluate the performance of the biometric measurements including dice score (Dice), absolute error (A-Error) and relative error (R-Error) of the circumferences from the prediction and the annotation. As shown in Table~\ref{tab:seg}, the proposed method gained good performance in Dice score. Meanwhile, the absolute error and relative error of fetal head circumference and abdominal circumference of our method are 1.125mm, 2.05$\%$ and 3.608mm, 3.25$\%$, respectively. The p-values in Table~\ref{tab:seg} also indicate our predicted biometrics has no significant difference with the annotations. This shows a similar performance with human-level performance~\cite{wu2017cascaded,van2018automated} and suggests that the proposed method has potential to be applied in real clinical setting.
	
	\subsection{Qualitative evaluation}
\begin{figure*}[htbp]
		\centering
		\includegraphics[width=0.95\linewidth]{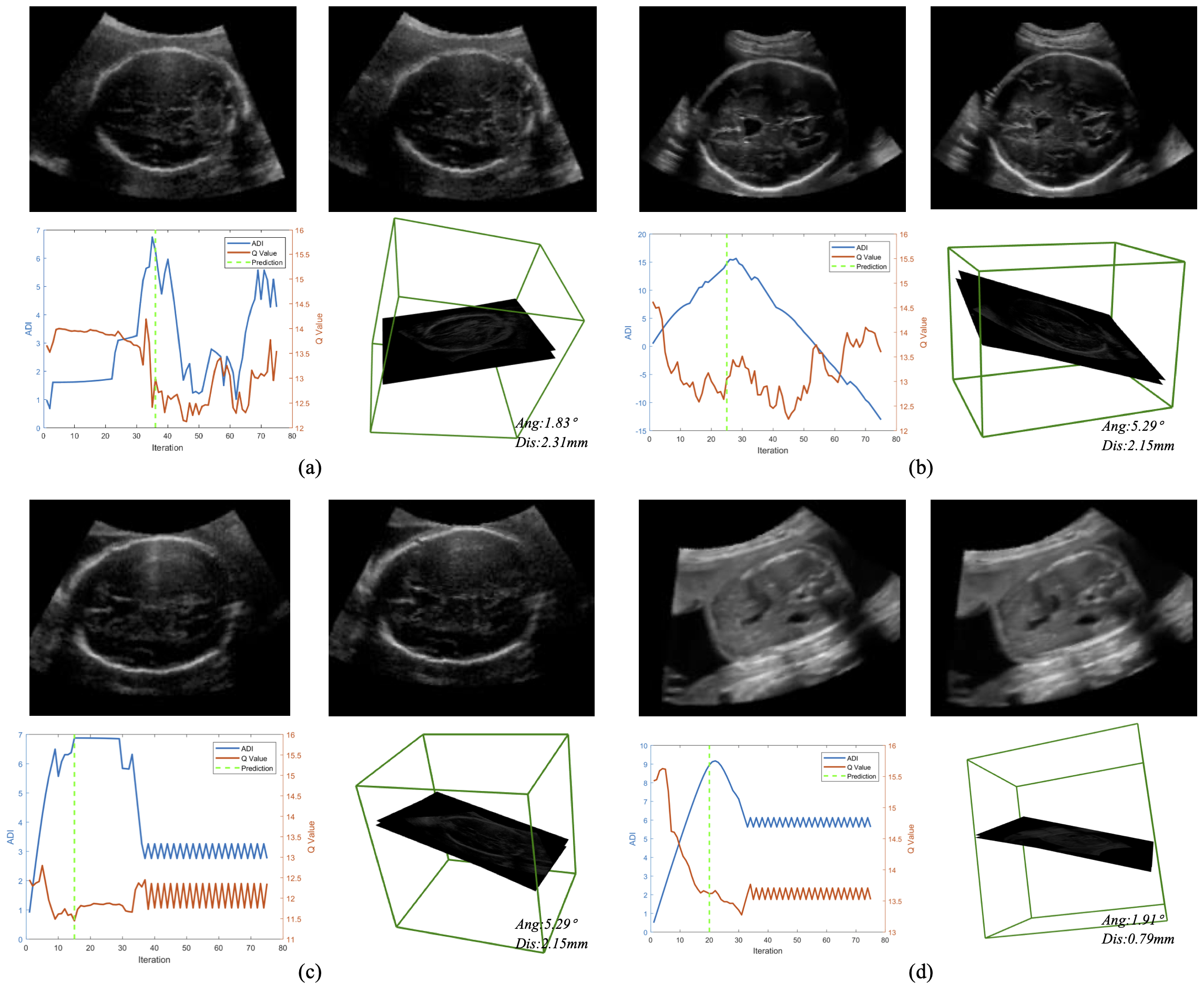}	
		\caption{Visualization of our method in sampled SPs of the US fetal dataset. (a) is the transcerebellar SP, (b) is the transventricular SP, (c) is the transthalamic SP, and (d) is the abdominal SP. For each case, the upper left is the predicted standard plane, the upper right is the ground truth, the bottom left is the inferring curve of the termination module, and the bottom right is the 3D spatial position of the predicted plane and ground truth.}
		\label{fig:visualization_fetal}
	\end{figure*}

\begin{figure*}[htbp]
		\centering
		\includegraphics[width=0.95\linewidth]{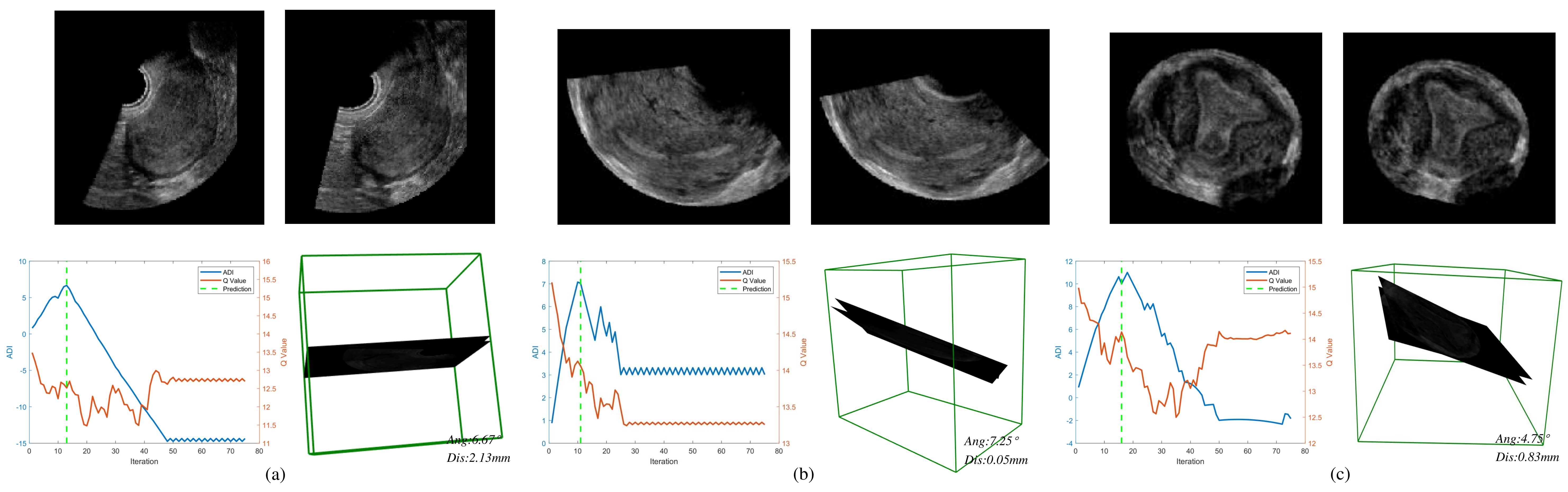}	
		\caption{Visualization of our method in sampled SPs of the US uterus dataset. (a) is the mid-sagittal SP, (b) is the transverse SP, and (c) is the coronal SP. For each case, the upper left is the predicted standard plane, the upper right is the ground truth, the bottom left is the inferring curve of the termination module, and the bottom right is the 3D spatial position of the predicted plane and ground truth.}
		\label{fig:visualization_uterus}
	\end{figure*}

	Figure~\ref{fig:visualization_fetal} and~\ref{fig:visualization_uterus} provide visualization results of the proposed method. It shows the prediction plane, the ground truth, the termination curve and the 3D spatial visualization of four randomly selected cases. It can be observed that the predictions are spatially close and visually similar to the ground truth. Furthermore, the proposed method can reach an ideal stopping point consistently. Both the maximum iteration and lowest Q values termination strategies fail in spotting the optimal termination step. 
	
	%========================== Discussian ==============================%
	\section{Discussions}
	Although RL is powerful in localizing view plane in MRI~\cite{Alansary10.1007/978-3-030-00928-1_32}, it failed to localize SPs localization in 3D US. Without an alignment module and early-stop setup, the \textit{AVP} needs a careful design for agent training and inference in a vast search space. Thus it is easier for learning-based localization methods to locate the SP within a limited search space. This might explain the relative low performance of ~\cite{Alansary10.1007/978-3-030-00928-1_32} in Table~\ref{tab:warm_up_fetal} and Table~\ref{tab:warm_up_uterus}. The proposed landmark-aware alignment module was devised based on the exact concern. It aligns all the volumes to the same atlas space using rigid registration, which can constrain the environment like that in MRI images. Furthermore, our proposed alignment method can be regarded as a prior-based initialization of the agent in testing US volumes, which reduces the search space into a fine-grained subspace.
	
	A proper termination strategy is essential in deep RL while it is difficult to estimate the optimal termination step because the agent often gets trapped in the local minimum during the iterative searching process. Prior studies have proposed several different termination strategies for such applications~\cite{Ghesu2019MultiScaleDR, Alansary10.1007/978-3-030-00928-1_32}. However, as shown in Table~\ref{tab:termination_fetal} and~\ref{tab:termination_uterus}, Fig.~\ref{fig:visualization_fetal}, and~\ref{fig:visualization_uterus}, the aforementioned experimental or previous knowledge-based termination strategies failed in estimating the optimal termination step in this challenging task. Meanwhile, the prior studies~\cite{Alansary10.1007/978-3-030-00928-1_32, Dou10.1007/978-3-030-32254-0_33} default the agent terminates at the fixed maximum step, causing inefficiency of the localization system. Our previous study designed a learning-based active termination using RNN to learn the mapping between the Q-value sequence and the optimal step. However, it requires waiting for the agent to finish inference as well. In contrast, our termination module enables the dynamic agent searching with the RNN to learn the implicit relationship between the Q-value curve and the optimal termination step. The resulting RL framework can achieve more accurate efficient predictions. Note that this learning-based termination strategy is a general method and can be applied to other similar tasks.
	
	%========================== Conclusion ==============================%
	\section{conclusion}
	In this paper, we present a deep RL framework equipped with 1) a landmark-aware alignment module to provide a warm start for the agent searching, and 2) a novel learning-based strategy for adaptive dynamic termination. SP localization in 3D US is challenging due to the low image quality, large data size and diverse fetal postures. Along with the proposed landmark-aware alignment module, the deep RL framework can start searching within the environment constrained by anatomical prior knowledge. In reinforcement learning for SPs localization, the termination conditions are usually indistinct and can not be precisely defined. Our proposed adaptive dynamic termination raises a new solution towards an effectiveness- and efficiency-steered localization system. Validation experiments showed that our model not only outperforms the current state-of-the-art learning based methods in detecting SPs, but also saves about 67\% time during inference and shows great generalizability across multiple challenging datasets.

	\ifCLASSOPTIONcaptionsoff
	\newpage
	\fi
	
	\bibliographystyle{ieeetr}
	\bibliography{references}
	
\end{document}